\documentclass{icrc29}
\usepackage[english]{babel}
\usepackage[latin1]{inputenc}
\usepackage{times}
\usepackage[T1]{fontenc}
\usepackage{hyperref}
\usepackage{epsfig}

\usepackage{graphicx}
\usepackage{amssymb}

\hypersetup{
pdfauthor={Pierre Billoir, Antoine Letessier-Selvon, CBPF - In2p3/CNRS},
pdfborder={0 0 0},
pdfview={Fit}
}
		  
\begin{document}
\title{A coverage independent method to analyze large scale anisotropies}

\author[Billoir, Letessier-Selvon]%
{Pierre Billoir \& Antoine Letessier-Selvon
\email{LPNHE - IN2P3/CNRS, 4 place Jussieu, 75252 Paris, France. (billoir@lpnhep.in2p3.fr),\\
CBPF-IN2P3/CNRS. Rua Xavier Sigaud 150, Rio de Janeiro, Brazil. (Antoine.Letessier-Selvon@in2p3.fr)}
}

\presenter{}
\maketitle

\begin{abstract}
The arrival time distribution of  cosmic ray events is well suited  to extract
information regarding sky anisotropies.  For an experiment with nearly constant exposure, 
the frequency resolution one can achieve is given by the inverse of 
the time $T$ during which the data was recorded. For $T$ larger than one calendar year the resolution becomes 
sufficient to resolve the sidereal and diurnal  frequencies.  Using a Fourier expansion on a modified time parameter, we show in this note that one can accurately extract sidereal modulations without knowledge of the experimental coverage. This procedure also gives the full frequency pattern of the event sample under studies which contains important information about possible systematics entering in the sidereal analysis. We also show how this method allows
to correct  for those  systematics.
Finally, we  show that a two-dimensional analysis, in the form of the spherical harmonic ($Y_l^m$) decomposition, can be performed under the same conditions for all $m\ne 0$. 
\end{abstract}

\section[Introduction]{Introduction and Notations}
 If anisotropies of the incident cosmic ray flux depend on right ascension, they induce on the observed event rate a
modulation with a period equal to one sidereal day. If other sources of modulations (weather effects, 
detector acceptance, etc), be them known or unknown to the experimenter, exhibit some periodicities, they have no 
reasons to be related to the sidereal period. In general, the extraction of large scale anisotropy rely on the difficult task of computing the detector coverage map, i.e. the detector 
response to a perfectly uniform sky. 
This estimation is often delicate as unknown or ill-corrected effects  may introduce systematics 
in this map, limiting the precision on the genuine anisotropies from a given data set.

The Fourier transform over an exposure time $T$ allows to separate frequencies with a resolution of the order of $1/T$.  
Data acquired over a few years,  even with variable detection conditions, can be analyzed using this transformation  allowing for a clean separation between  the sidereal frequency and the diurnal frequency,  which differ by  1 calendar day over a calendar year. This  separation can be obtained totally independently of the knowledge of the detector coverage map, hence avoiding the  difficulty of its estimation. At higher frequencies, when looking at multiples of the diurnal or sidereal frequency the resolution  is even better since the separation between the $n^{th}$ harmonics is $n$ times larger than between the fundamentals .

The Fourier transformation is an ideal tool to extract the frequency patterns from a set of event times. We define the (normalized)  Fourier coefficients  $a(f)$ of set of times \{$t_i$\} at frequency $f$ (cycle $\omega = 2\pi f$) as :

\begin{eqnarray}
a(f) = c(f) + is(f)
\end{eqnarray}
with  
\begin{eqnarray}
c(f) = \frac{2}{W}\sum_i w_i\cos{\omega t_i}\\
s(f)  = \frac{2}{W}\sum_i w_i\sin{\omega t_i}
\end{eqnarray}
with $W=\sum_i w_i$ and where the sums run over all the $N$ events whose time distribution is studied.
The event weights $w_i$ may incorporate known modulations of the data that one wants to  remove from the 
expansion.  When no correction is desired $w_i=1\; \forall i$, and $W = N$\footnote{We will place ourselve in this case except in section~\ref{1d-ex}.}.
The amplitude at frequency $f$ is then given by $|a(f)| = \sqrt{s^2(f)+c^2(f)}$ while the phase is defined as 
$\phi(f)  = \tan^{-1}{(s(f),c(f))}$.

In general cosmic ray surface detectors have a large instantaneous aperture and can observe a sizeable fraction 
of the sky at any given time. For example, a surface array whose detector can see air showers with zenith up to 60 degrees
sees instantly 1/4th of the whole sky. To be able to observe modulations on smaller scale it is therefore necessary to correct the 
time entering in Eqn. 1 and 2 by the right ascension phase of the event with respect to the Local Sidereal Time (LST, 
the right ascension of the stars above the local meridian).   We then modify $t_i$ entering in the equations above as:
\begin{eqnarray}
{\xi_i} = t_i + (RA_i - LST(t_i)) 
\end{eqnarray}

where $LST(t_i)$ is the local sidereal time corresponding to UTC time $t_i$ 
and $RA_i$ is the right ascension of the event. Note that $LST({\xi_i}) = RA_i$. 
With such a definition the Fourier Expansion in Modified Time (FEMT) (with unit weights $w_i=1$) 
is equivalent at the sidereal frequency  to the Rayleigh
analysis. Note that without the introduction of 
this modified time an eventual high order (small angular scale with respect to the instantaneous aperture) 
modulation in  sidereal time would average almost to zero. On the other hand an eventual diurnal modulation 
will be slightly reduced when using this parameter.

\section{Properties}
In this section we show how the  Fourier expansion allows to completely separate, at first order, sidereal time modulations from experimental modulations without the need to correct for  the detector coverage map.

A given event population $P(d,{\xi})$ with declination $d$ is proportional to :
\begin{eqnarray}
P(d,{\xi}) = k F(LST({\xi}),d) \int_{t,\theta,\varphi} A(\theta, \varphi, t) \delta({\xi} -t) \delta(d - d(\theta,\varphi) dtd \cos{\theta}d\varphi
\end{eqnarray}
i.e.
\begin{eqnarray}
P(d,{\xi}) =  k F(LST({\xi}),d) \sum_{n=-\infty}^{+\infty} A'(d,t+nT_{sid})= k' F(ra,d) A_s(d,t)
\end{eqnarray}
where $k,k'$ are constants, $F$ represents the sky flux in the direction $(ra,d)$ and $A$ ($A'$, $A_s$) is the detector acceptance 
corresponding to our event selection and $T_{sid}$ is the sidereal period. 

The Fourier transform, $\tilde{P}$ of $P$ is the convolution product of the Fourier transform $\tilde{F}$ of F and $\tilde{A}_s$
of $A_s$ :

\begin{eqnarray}
\tilde{P}(d,f) = K \int df'\tilde{F}(d,f') \tilde{A}_s(d,f-f')
\end{eqnarray} 

Because $F$ represents the sky flux it is a perfectly periodic functions of time (or of $\xi$) and $\tilde{F}$ is given by an harmonic series of the fundamental sidereal frequency $f_{sid} = 1/T_{sid}$ :

\begin{eqnarray}
\tilde{F}(d,f) = \sum_m a_m(d)\delta(f-mf_{sid})
\end{eqnarray} 
with $a_0=2$ due to our definition of the Fourier transformation above. 

To simplify the notation we will also assume that  $\tilde{A}_s$ is an harmonic series but of the diurnal frequency $f_d$ :
\begin{eqnarray}
\tilde{A}(d,f) = \sum_m b_m(d)\delta(f-mf_d)
\end{eqnarray}
again with $b_0 = 2$. 
$\tilde A$ may contain other fundamental frequencies such as $f_y$ and also be non periodic.\footnote{A detector only active during a period of time $T$ is a square function and its Fourier transform would be the convolution of $1/(2\pi f T) \times \sin{(2\pi fT)}$ with the harmonic development above.} The one important assumption we make is that $f_{sid}$ does not enter as a special  frequency in $\tilde A$ as might  $f_d$ or $f_y$.

We can assume that the modulations introduced by $a_m$ and $b_m$ with $m\neq 0$ are small (the acceptance is relatively stable in time and the true anisotropies, if any, are small).  Then, at first order in $a_m$ and $b_m$, $\tilde{P}$ reduces to:
\begin{eqnarray}
\tilde{P}(d,f) \simeq 2 + \sum_{m>0}a_m(d)\delta(f-mf_{sid}) + \sum_{m>0} b_m(d)\delta(f-mf_d)
\end{eqnarray}
We observe  a decoupling of the modes.  The relative modulations at frequency $f$ in each declination band $d$ 
is directly given by the modulus of $\tilde{P}$. In particular for each sidereal harmonic $m$ we have :
\begin{eqnarray}
|\tilde{P}(d.mf_{sid})| =  |\tilde{P}(d.m)| \simeq \sqrt{c^2(mf_{sid})+s^2(mf_{sid})}
\end{eqnarray}
Here $c$ and $s$ are given by eqn. 2 and 3 above but with the sum running over all events with declination $d$.

Moreover, the knowledge of the evolution of $\tilde{P}(d,m)$ as a function of the declination $d$ 
allows for the reconstruction of the corresponding multi-polar coefficients under certain conditions; if $F$ is expressed as:
\begin{eqnarray}
F_0\left(1+\sum_{(l,m)\neq(0,0)} \alpha_{lm} Y_l^m(\pi/2-d,ra)\right)
\end{eqnarray}
the function $a_m(d)$, for a given value of $m$, gives 
$\sum_{l\geq m} \alpha_{lm} {\cal P}_l^m(\pi/2-d)$, where the ${\cal P}_l^m$ 
are the associated Legendre polynomials. If the expansion may be truncated to a certain order, we can deduce from these constraints the decomposition of F over the $Y^m_l$ ($m\neq 0$), independently of the knowledge of $A$ (but the precision is limited by the available statistics, of course).

In practice experimental effects will of course partly blur this ideal picture, in particular :
\begin{itemize}
\item The sum in eqn. 2 and 3 only runs over a finite amount of time T hence the 
Fourier transform is convoluted with the transform of the function describing the acceptance as a function of the time. 
With a constant acceptance it is a function ($\sin{x}/{x}$) with $x=\omega T$; it is more complex
when accounting for possible overall increase or short scale irregularities of the detector up time, but in any case, with $T$ larger than a few years it has a width less than  the difference between the 
sidereal and diurnal frequencies. We will see below that this resolution is sufficient to perfectly separate the two components.
\item We have a finite number of events N, hence a Poisson noise in all frequencies below about N/T, this will induce an average statistical noise on our amplitude estimates of order $\sqrt{\pi/N}$.
\item The diurnal modulation are not perfectly periodic with a 24 hour period. Moreover annual variation over a period of 
about a year  ($355.25*f_d$) also exist. This may produce second order (the amplitude is the product of the annual and diurnal amplitude) side bands at $f_{d}+f_{y} \simeq f_{sid}$ and 
$f_{d}-f_{y} \simeq f_{anti-sid}$. 
In practice the  existence of such bands can be checked on the anti-sidereal (lower frequencies) side of the frequency 
distribution. Therefore, after a first order reconstruction of $\tilde{F}$ and $\tilde{A_s}$ from eqn. 10, one 
can evaluate the contribution  of the second order terms, and,  if necessary, to re-evaluate $\tilde{F}$ correcting for the principal modulations of $\tilde{A_s}$. An example of such corrections is given in section~\ref{1d-ex}. 
\end{itemize}

\section{Noise statistics and upper limits}
In the case of a totally uniform sky and in the absence of local effect the time entering in equation 1-3 are uniform and the variable $c(f)$ and $s(f)$ converge, in virtue of the central limit theorem, toward Gaussian random variables
 with mean $0$ and variance $2/N$. Due to the nature of $c(f)$ and $s(f)$ the convergence is very rapid and holds in practice for $N$  of order a few tens. The random variable $z(f) = \sqrt{c(f)^2+s(f)^2}$ is then a Rayleigh random variable
 of parameter $\sigma=\sqrt{2/N}$ whose distribution is :
 
 \begin{eqnarray}
 f(z\;\mathbf{|}\;\sqrt{2/N}) = \frac{Nz}{2}e^{-Nz^2/4}
 \label{rayleigh}
 \end{eqnarray}
 
 The mean of this distribution is $\sqrt{\frac{\pi}{N}}$ and the variance is $\frac{4-\pi}{N}$.  The cumulative distribution function
 giving the probability for the noise to be above some power $z_0$ is given by :

  \begin{eqnarray}
 F(z_0\;\mathbf{|}\;\sqrt{2/N}) = 1 - e^{-Nz_0^2/4}
 \end{eqnarray}

Hence, as in the standard Rayleigh analysis of cosmic ray data,  for a background sample of size $N$ the $\alpha$ CL upper limit on the noise power is given by $p_{\alpha}=-2\sqrt{\ln{(1-\alpha)}/N}$. 

In practice what is often needed is, given a measured power $p$,  the probability to observe this measurement or less  when a non random modulation of power $p_0$ is present in the data set.  When signal is present the distribution of $z$ is no longer given by Eqn.~\ref{rayleigh}  but 
by the Rice distribution : 

\begin{eqnarray}
R(z\;\mathbf{|}\;\sqrt{2/N}, p_0) = \frac{Nz}{2}e^{-N(z^2+p_0^2)/4}I_0(zp_0N/2)
\end{eqnarray}

where $I_0(x) $ is the modified Bessel function. 

The $\alpha$ confidence level upper limit $p_\alpha(f)$ on a signal at frequency $f$ when a power $p$ has been observed can be defined as the solution of:

\begin{eqnarray}
\int_{p}^{+\infty} R(z\;\mathbf{|}\;\sqrt{2/N}, p_\alpha) dz = \alpha
\end{eqnarray}
which can be integrated numerically.

When the systematical noise, such as the resolution noise or the side-band noise, becomes dominant over the statistical noise the distribution of the reconstructed power at the sidereal frequency is only a Rice distribution for a fixed signal phase with respect to the systematical noise.  When the signal phase is random the power distribution becomes a superposition or Rice distribution with different $p_0$ parameters (depending on the relative phase between the systematical noise and  the signal\footnote{For a signal $p$ with a uniform random phase and a systematical noise of amplitude b with phase $0$ the distribution of the Rice parameter $p_0$ becomes $f(p_0) \propto  \sqrt{p^2+b^2+2pb\cos{\theta}}/(2pb\sin{\theta})$ with $\theta$ uniform in $[0,2\pi]$.}).

\section{One dimensional examples}
\label{1d-ex}
To illustrate the method and the various properties depicted in the previous section we constructed random data sets according to 3 different situations. All data sets have an annual modulation of 20\% with a fixed phase $\phi_y=1.0$, 
a diurnal modulation of amplitude 5\% and phase $\phi_d = 0.0$ and eventually either a dipole modulation of amplitude 1\% declination $0^\circ$ and phase $\pi/2$ or a right ascension modulation of amplitude 1.5\% and phase $\pi/2$. 

Data set {\bf A} contains 100,000 events and spans a period of 3 years, data set {\bf B} contains 400,000 events and
spans 5 years, finally data set {\bf C} contains 2,000,000 events and spans 15 years. Events are generated uniformly in UTC time and azimuth and with a zenith distribution proportional to $\sin{\theta}\times \cos{\theta}$ corresponding to a cosmic ray surface array located at longitude 69.25 West and latitude 35.25 South. The events are then accepted using an acceptance/rejection method according to the desired combination of modulations.  In table~1 a summary of the parameters of those various data sets is given.

\begin{figure}[tbp]
\begin{center}
\includegraphics*[width=0.85\textwidth]{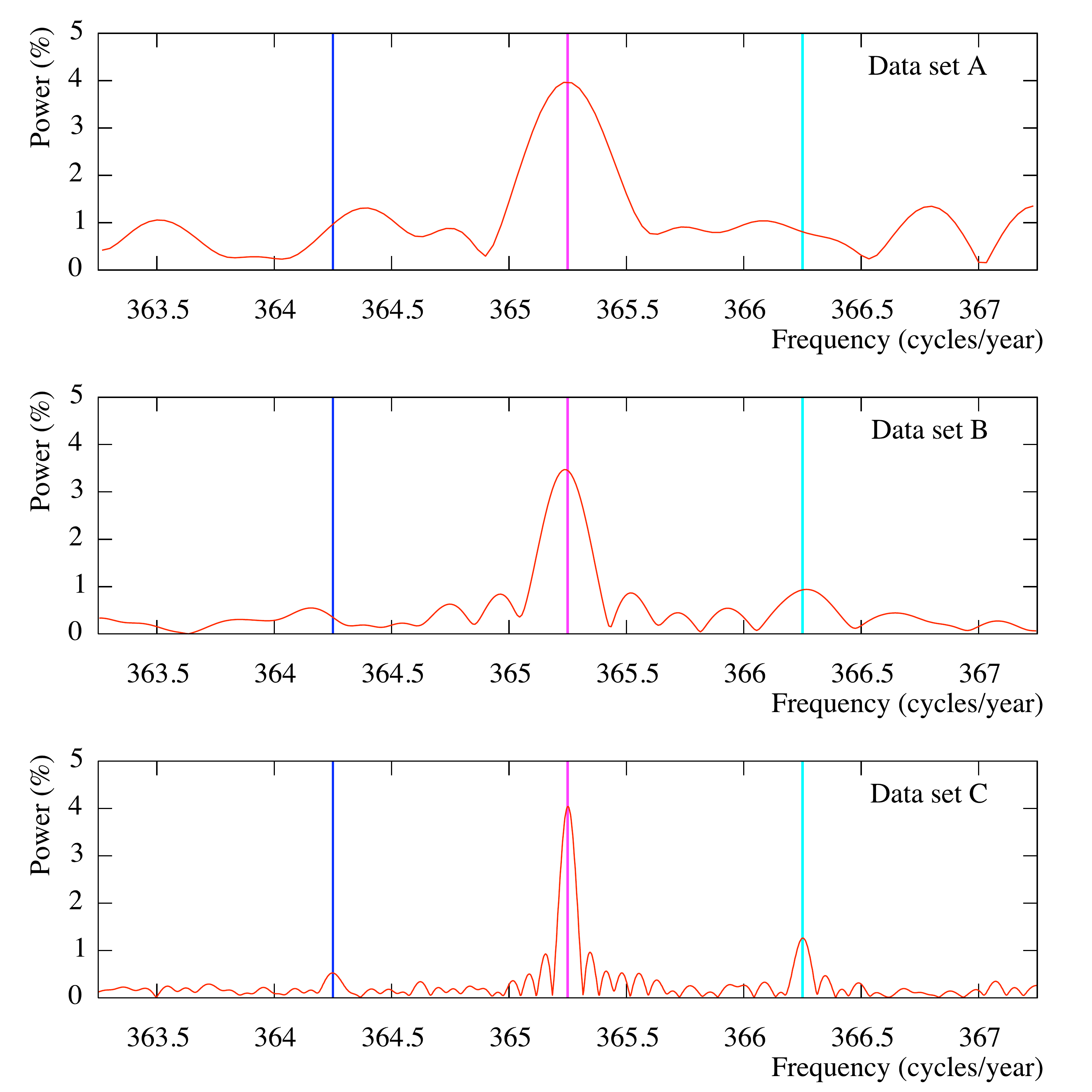}
\end{center}
\caption{Data sets {\bf A, B} and {\bf C} including the 1\% dipolar modulation (see text). Top : Fourier analysis of data set {\bf A}  around the diurnal modulation (central violet vertical line) . No signal can be seen at the sidereal (right vertical line) nor at the anti-sidereal (left vertical line) frequencies. Middle : Same for data set {\bf B}, the sidereal modulation is visible. Bottom : Data set {\bf C}, both the sidereal and anti-sidereal modulations are visible.}
\label{ABC-1}
\end{figure}

\begin{figure}[htbp]
\includegraphics*[width=0.33\textwidth]{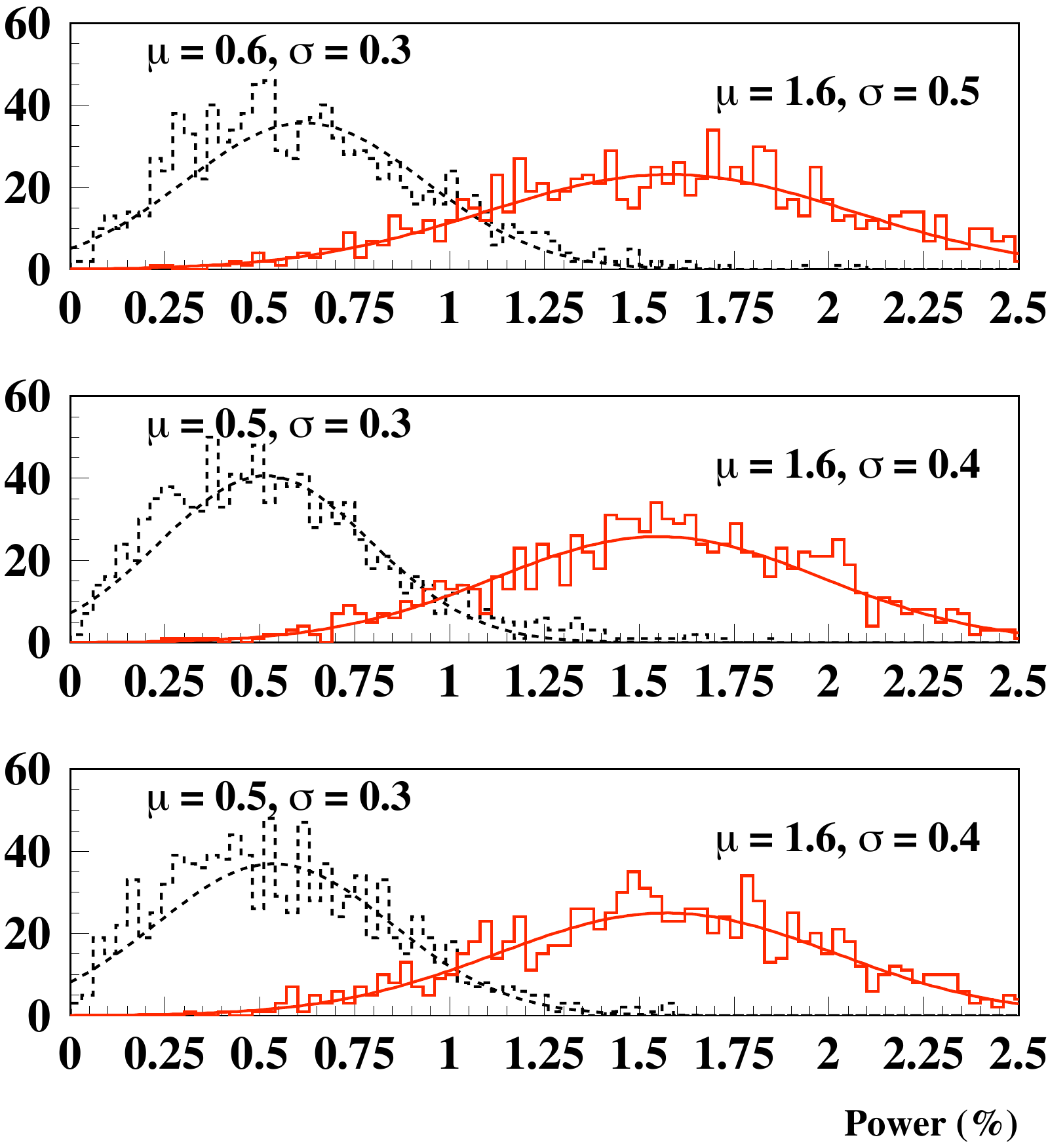}
\includegraphics*[width=0.33\textwidth]{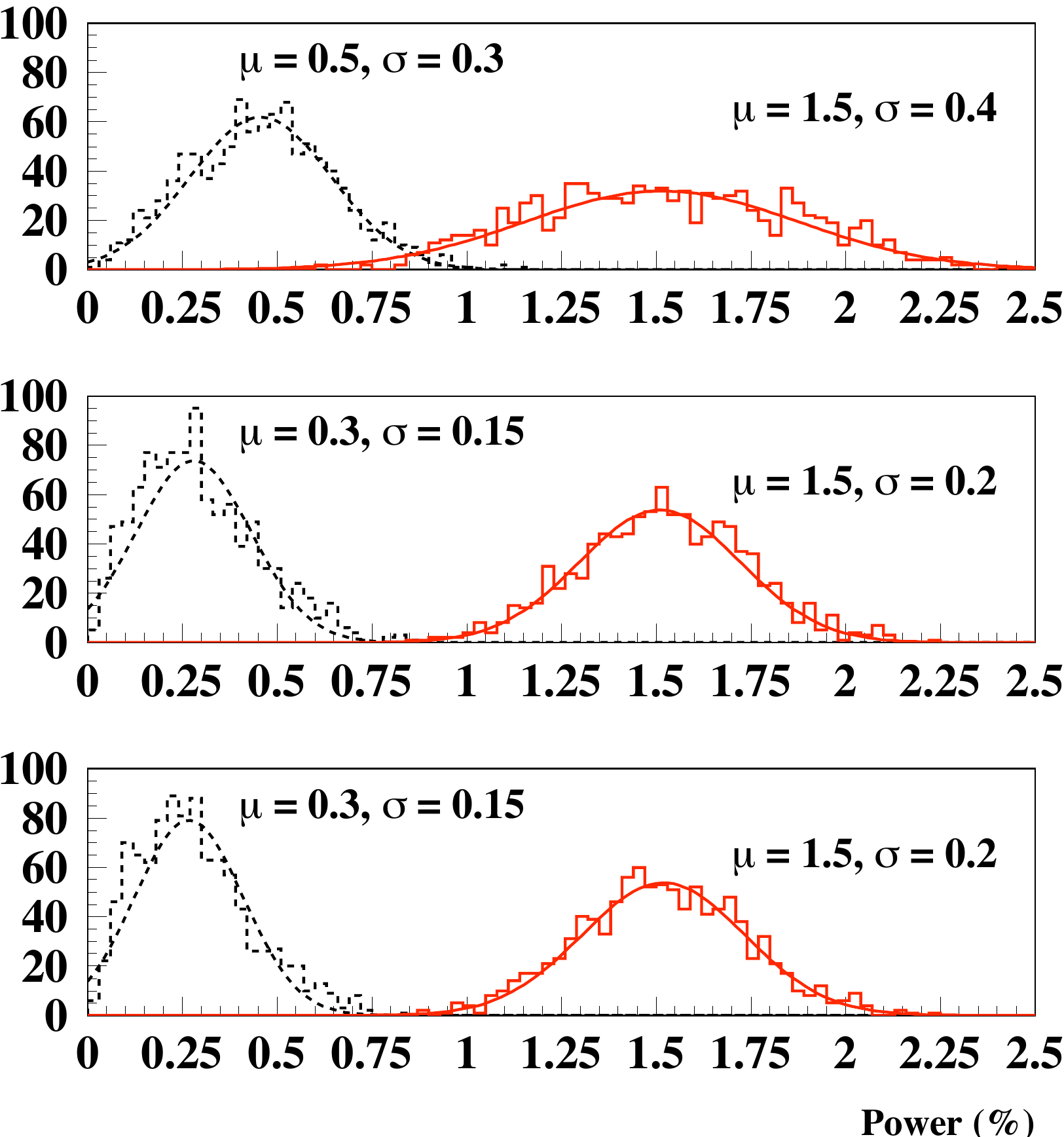}
\includegraphics*[width=0.33\textwidth]{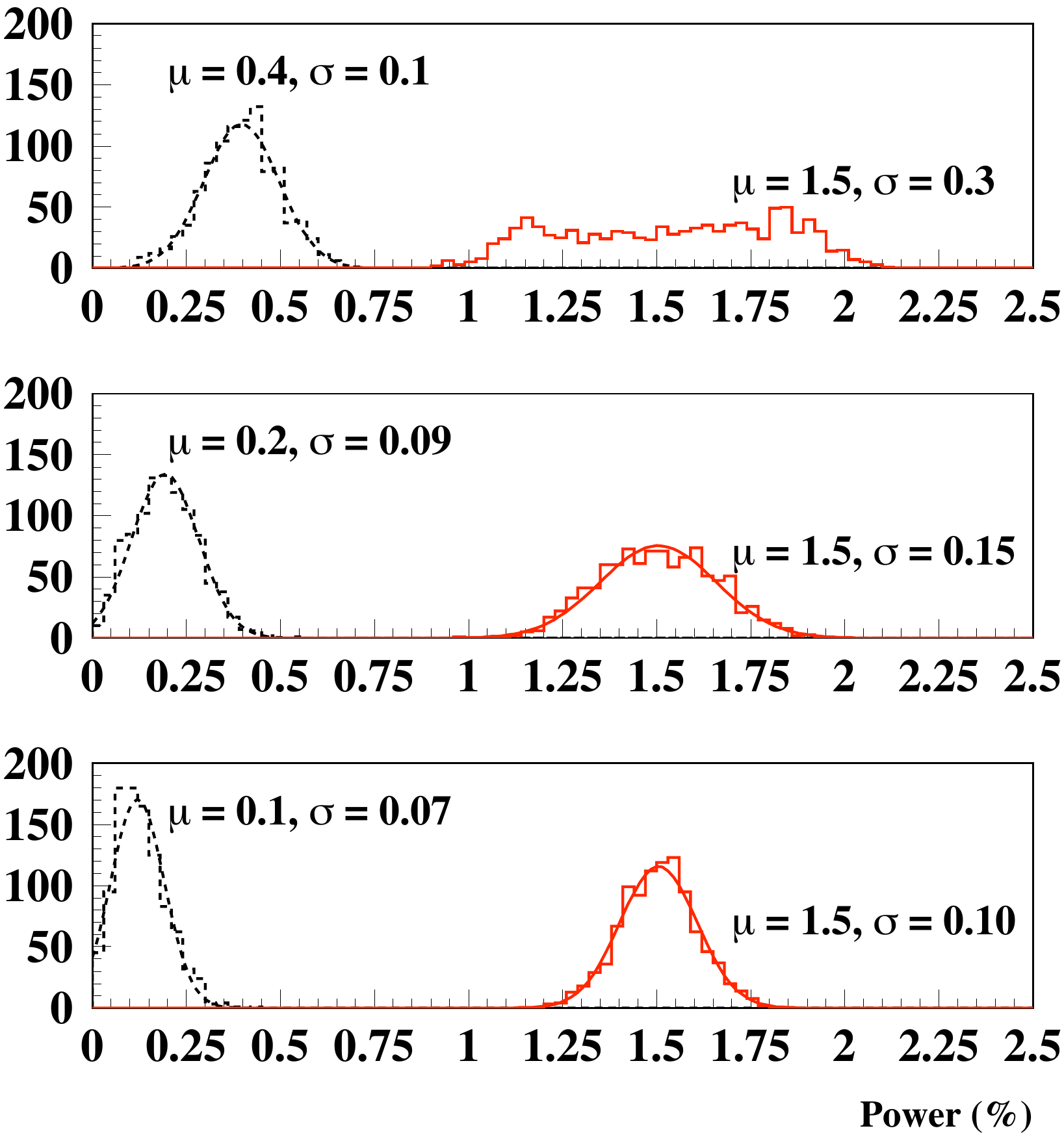}
\caption{Left : Power distribution at the sidereal frequency for data set {\bf A} with no sidereal modulation (dashed black) 
and with the 1.5\% RA modulation (solid red).The curves are gaussian fit to the distributions with the parameters indicated.  The top histogram corresponds to the Fourier transform on the
original data set, the middle histogram includes the correction for the annual modulation (see text) and the bottom one for both annual an diurnal. Since the statistical noise is dominant in data set {\bf A} the improvement is marginal after the correction. Middle : Same for data set {\bf B}. Now the improvement is clearly visible between the top histogram and the two lower ones.  Right : Same for data set {\bf C}. Again the improvement is clearly visible, the separation between the average noise and the 1.5\% modulation increase from top to bottom from 3.7$\sigma$ to 14$\sigma$. Note that in this case the red curve of the top histogram which corresponds to the power distribution of a 1.5\% modulation is no longer  a Rice distribution because we are in this case dominated by the side-band noise which has a fixed phase. }
\label{ABC-2}
\end{figure}

\begin{table}[tdp]
\label{tab1}
\begin{center}
\begin{tabular}{c|c|c|c|c|c|c|c}
Data set & N & $p_y $ & $p_d$  & $\mu_{stat}$  & $\sigma_{stat}$  & side band  & diurnal resolution noise  \\ 
-& -&- & -& ($\sqrt{\pi/N}$) & ($\sqrt{(4-\pi)/N}$) & $(p_y\times p_d)/2$ & $\sim p_d/(\pi T)$ (T in year).\\ \hline
{\bf A} & $10^5$ & 19 & 4 & 0.6 & 0.3 & 0.5 & 0.4 \\ \hline
{\bf B} & $4\times10^5$ & 19 & 4 & 0.3 & 0.15 & 0.4 & 0.3 \\ \hline
{\bf C} &$2\times 10^6$ & 19 & 4 &  0.1 & 0.07 & 0.4 & 0.1 \\ 
\end{tabular}
\caption{Background parameters of the different data sets. In data set {\bf A} the statistical noise dominates, in data set {\bf B} statistical and systematical noise are similar, in data set {\bf C} the systematics dominate. Note that the amplitude
given for the annual ($p_y$) and diurnal ($p_d$) modulation are the one reconstructed in modified time hence they are smaller than the original ones in UTC time (20\% and 5\% respectively). The diurnal resolution noise is given at the sidereal (and anti-sidereal) frequency. All numbers, except $N$, are in percent.}
\end{center}
\end{table}%

On figure~\ref{ABC-1} the FEMT  power as a function of frequency is shown around the diurnal frequency
for data set {\bf A, B} and {\bf C} including the 1\% dipolar modulation.  For data set {\bf A} only the diurnal modulation is visible with a reduced amplitude of 4\% due to the use of the modified time. Neither the sidereal modulation nor the anti-sidereal side-band are visible. The resolution, given by the width of the diurnal peak,  is however sufficient to resolve the diurnal and sidereal frequency. With data set {\bf B}  the situation is clearer, the improved statistics and resolution now allow to see the sidereal modulation, the anti-sidereal band is still invisible. Finally with data set {\bf C} one can perfectly resolve the 3 modulations. Note that the anti-sidereal band has an amplitude of about 0.5\% as expected ($\sim 0.2\times0.05/2$).

On figure~\ref{ABC-2} we show the power distribution at the sidereal frequency obtained from 1000 random sets of type {\bf A, B} or {\bf C} (from left to right). The first row of histograms shows the FEMT with unit weights. Dashed (black) curves correspond to the case
without RA modulation while solid (red) ones correspond to the case with the 1.5\% RA modulation. One clearly sees on 
data set {\bf C} the effect of the 0.5\% systematic modulation which dominate over the noise and gives a power distribution (solid red) almost uniform between 1 and 2\%. 

On the second row of histogram the FEMT is evaluated using event weights. For each event with time $t_i$ the weight  $w_i$ is given by :
\begin{eqnarray}
w_i =  \left [ 1+p_y\cos{(\omega_y t_i -\phi_y)} \right ] ^{-1}
 \end{eqnarray}
 where $p_y$ and  $\phi_y$ are the power and phase measured at the 
 annual frequency ($\omega_y$) using a unit weight FEMT. 
 As expected the improvement is only marginal for data set {\bf A} but very significant  for data sets {\bf B} and {\bf C}. 

Finally on the last row of histograms of figure~\ref{ABC-2} we show the power distribution using a weighted FEMT
correcting for both the annual and diurnal modulations :
 \begin{eqnarray}
 w_i = \left [ (1+p_y\cos{(\omega_y t_i -\phi_y)}\times(1+p_d\cos{(\omega_d t_i -\phi_d)} \right  ] ^{-1}
 \end{eqnarray}
 
where $p_y$ and  $\phi_y$ are as above and $p_d$ and $\phi_d$ are the power and phase measured at the 
 diurnal frequency ($\omega_d$).  Again the improvement is significant for data set {\bf C}. It is remarkable to note 
 that after this correction all data sets show a background power distributions corresponding exactly to the expectation
 from their limited statistics. 

\begin{figure}[tbp]
\includegraphics*[width=0.50\textwidth]{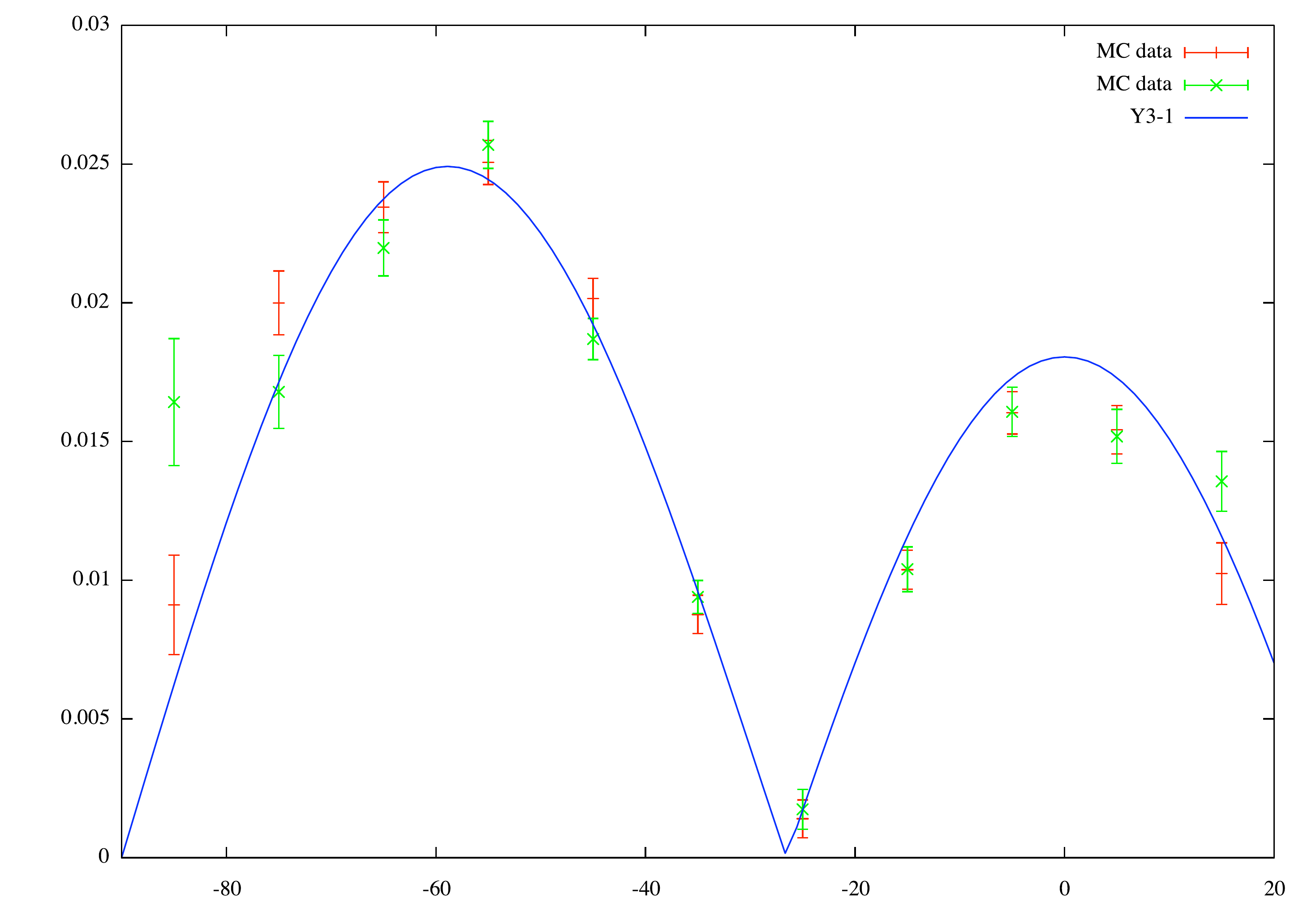}
\includegraphics*[width=0.50\textwidth]{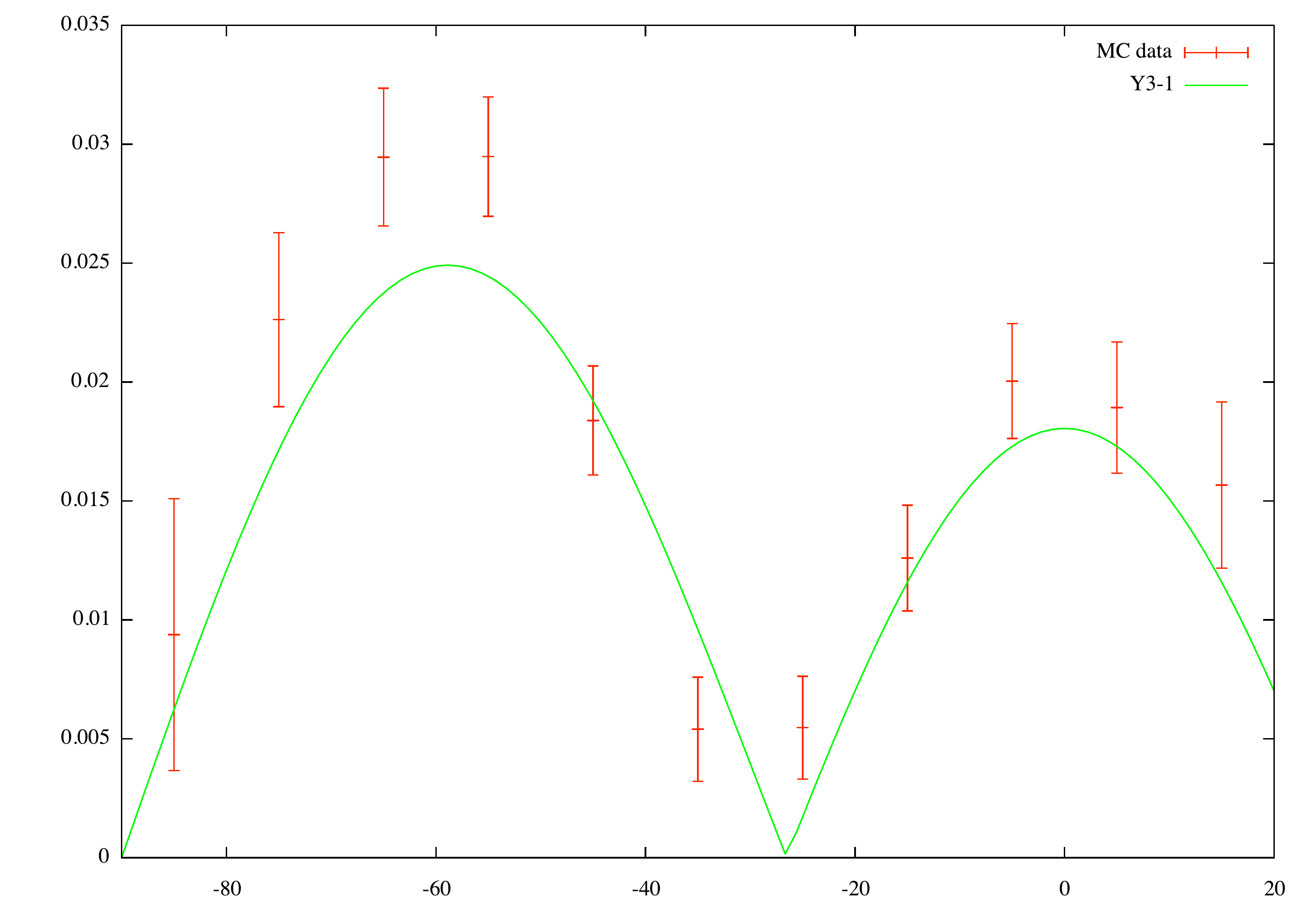}
\caption{Left : Reconstructed modulation at the sidereal frequency as a function of declination, for a sky with a 5\%  $Y^3_1$ modulation and 
$N=4\times 10^{6}$. 
In green  uniform coverage in zenith, in red coverage (proportional to $\cos{\theta}\sin{\theta}$). 
The (blue) curve represents the genuine input modulation. As can be seen, although not corrected for,
 the coverage does not influence the reconstructed amplitude evolution with declination. 
 Right a typical plot for $400,000$ events, a 5\% $Y^3_1$ modulation and a coverage proportional to $\cos{\theta}\sin{\theta}$. }
\label{Ylm}
\end{figure}

\section{Bi-dimensional application}
The analysis depicted in the previous section dealing with a single declination band is quite appropriate to analyze dipole moments ($l=1$). However,  the power given by equation~11 can be derived, if the statistic is sufficient, in several distinct declination bands allowing to reconstruct higher order moments. We just give a taste of the method using a sky  modulated 
by a 5\%  $l=3$, $m=1$ $Y_l^m$.  

On figure~\ref{Ylm} left, we show the reconstructed power for 11 bands in declination (from -90 to +20). 
A total of $4\times 10^{6}$ events were generated using two different zenith coverage : uniform coverage in zenith ($\theta$ in the range [0, 90] degrees) and a  coverage proportional to $\cos{\theta}\sin{\theta}$ in the range [0, 60] degrees). In both case the centre of the detector is at location 
(69.25$^\circ$ W, 35.25$^\circ$ S).
This  shows
 that our analysis results are, as predicted,  independent of the experimental coverage.  On the right figure we show
 the expected modulation that would be seen  if a 5\% $Y^3_1$ was present on the sky for a statistics of 400,000 events
 over 5 years.

The power distribution as a function of declination can be reconstructed independently of the coverage because Eqn~11 gives the fraction of events that are modulated within that declination band, independently of the relative normalization of one band to another.  Hence the result is independent of the eventual variation of the coverage with the declination.

\section{Conclusion}
We have constructed a modified Fourier transformation which allows to reconstruct the original right ascension  modulation of a data set without the need to construct the appropriate coverage map. We also showed that this technique allows to obtain, for each order $m \neq 0$ in right ascension, the dependence in declination over the range covered by the observations, independently of the coverage function. 

\end{document}